# On the existence of the *NN*-decoupled dibaryon $d_1^*$(1956).

**A.S. Khrykin**[1a]

[1]Joint Institute for Nuclear Research, Dubna, 141980 Russia

**Abstract.** We present strong experimental evidence for the existence of a nonstrange dibaryon with a mass of about 1956 MeV stable against strong decay, which is a very likely candidate for long-sought multiquark hadrons with the baryon number ***B*=2**. We start by presenting the first evidence for the existence of this dibaryon called *$d_1^*$(1956)*, that was found in the energy spectrum of coincident photons emitted at ±90⁰ from the reaction ***pp→ppγγ*** at an energy of 216 MeV measured at JINR. We then show its signatures found in several experimental energy spectra of single photons and invariant mass spectra of photon pairs from photon production processes induced by nucleon-nucleon and nucleon-nucleus collisions at intermediate energies.

## 1 INTRODUCTION

The problem of possible existence of multiquark hadrons with the baryon number *B*=2 (exotic dibaryons) was raised for the first time by R.L.Jaffe, who in the framework of the *MIT* model predicted existence of a dibaryon with strangeness *S*= **-2** stable against strong decay, called the *H* dibaryon [1]. Since then, this problem has been the subject of a great deal of theoretical and experimental efforts.

Unfortunately, static properties of exotic dibaryons (masses, widths and quantum numbers) are not calculable from the first principles of *QCD* since the relevant equations of the theory defy solutions by perturbation methods. At the same time, reliable calculations of these properties using the lattice *QCD* are still also impossible [2-4]. For this reason, most, if not all, predictions for the static properties of the exotic dibaryons have been obtained within the scope of *QCD*-inspired models [5-9]. Of course, the predictions obtained are highly model dependent. In this connection, the experimental discovery of any such hadron would have a big effect on our understanding of this theory and would provide valuable information for the stringent test of both *QCD* models and *QCD* itself.

Promising candidates for nonstrange exotic dibaryons would be *NN*-decoupled dibaryons with masses $M_R \leq 2m_N + m_\pi$ ($m_N$ and $m_\pi$ are the masses of the nucleon and the pion, respectively) [10,11]. These dibaryons should have the isospin *I*=2 or *I*=1(0) and such combinations of the total momentum *J* and parity *P* that are forbidden for *NN* systems by the Pauli Exclusion Principle: $I(J^P)$=**1(1⁺,3⁺,etc.)** or **0(0⁺,2⁻,4⁻,etc)**. It is clear that such *NN*-decoupled dibaryons should be stable against strong decay and could only decay into two nucleon states via electromagnetic interaction with the dominant decay mode $^2B \to NN\gamma$. It is natural to expect that they should be very narrow ($\Gamma_R \sim keV$ or even less).

If any *NN*-decoupled dibaryon with $M_R \leq 2m_N + m_\pi$ ($^2B$) exists in nature, it may be produced in the radiative capture process $NN \to {}^2B\gamma$. This process of the $^2B$ production would entail a new, dibaryon mechanism of the two-photon production in *NN* collisions $NN \to {}^2B\gamma \to NN\gamma\gamma$, which would compete with the conventional mechanisms of the reaction $NN \to NN\gamma\gamma$. In this regard, the reaction $pp \to pp\gamma\gamma$ possesses high sensitivity to the existence of such a mechanism, since its cross section due to conventional mechanisms is very small [12,13].

## 2. First Evidence for the $d_1^*$(1956) dibaryon.

The experimental evidence for the existence of the dibaryon mechanism of the two-photon production in *NN* collisions was found for the first time in the energy spectrum of coincident photons emitted at the laboratory angles **±90⁰** from the reaction $pp \to \gamma\gamma X$ at **216** MeV [14]. The spectrum is shown in Fig. 1 (left). It consists of a narrow peak at an energy of about **24** MeV and a relatively broad peak in the energy range (**45 - 75**) MeV. The statistical significances for the narrow ($S_N$) and broad ($S_B$) peaks are **5.3σ** and **3.5σ**, respectively. The width of the narrow peak (**FWHM**) was found to be about 9 MeV, comparable with that of the energy resolution of the experimental setup.

The analysis of this spectrum showed that its behavior conformed to the signature of the dibaryon mechanism $pp\to {}^2B\gamma \to pp\gamma\gamma$ of the reaction $pp \to pp\gamma\gamma$ with the ${}^2B$ mass ~ **1956** MeV. The obtained invariant mass distribution of the ${}^2B$ is shown in Fig. 1(right). The extracted mass of the dibaryon called $d_1^*(1956)$ is $M_R$ =1956 ±1 stat ±7 syst MeV. The energy spectrum of photons from the process $pp\to\gamma d_1^*(1956)\to pp\gamma\gamma$ calculated in the dipole approximation was found to be in good agreement with the measured one. Since the narrow and broad peaks of this spectrum refer to the signature of the $d_1^*(1956)$ its total statistical significance is $S_{tot} = (S_N^2+S_B^2)^{1/2} = 6.35\sigma$.

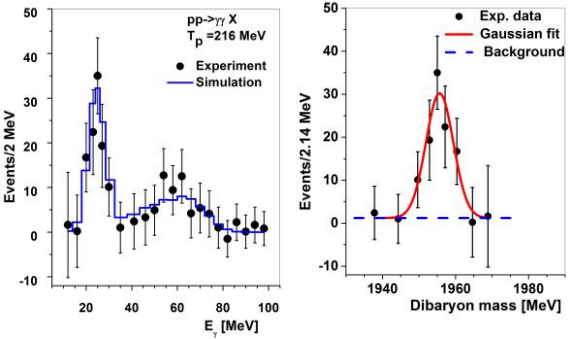

Fig.1 (Color online) Left: Experimental energy spectrum of photons emitted at $\pm 90^0$ from the process $pp\to\gamma\gamma X$ at 216 MeV [14] (full circles) compared to the Monte Carlo simulated energy spectrum of photons from the process $pp\to\gamma\ {}^2B \to\gamma\gamma pp$, for the dibaryon ${}^2B$ with a mass of **1956** MeV (solid line). Right: Experimental dibaryon mass distribution (full circles). The solid line indicates the Gaussian fit to the experimental data.

## 2. Indications for the $d_1^*(1956)$ in inclusive photon energy spectra.

Clear evidence for the existence of the $d_1^*(1956)$ dibaryon is demonstrated by the experimental inclusive energy spectra of photons emitted at the laboratory angle $90^0$ from the reactions $np\to\gamma X$ at **170±35** MeV [15,16], $pd\to\gamma X$ at **195** MeV [17] and $p{}^{12}C\to\gamma X$ at **200** MeV [18] shown in Fig.2. A common feature of these spectra is a structure in the energy range from **45** to **75** MeV, the origin of which is not explicable in terms of the conventional photon production mechanisms of these reactions [19]. At the same time, our calculations show that the mechanisms of photon emission associated with the $d_1^*(1956)$ production in these reactions give contributions to the same energy region where the structures reside and hence are very likely to cause their appearance. The presence of this structure in the spectrum of the $pd\to\gamma X$ reaction leads to serious disagreement between the experimental [17] and theoretical data [20]. In Fig. 2 we also show the calculated spectrum that corresponds to a sum of the $d_1^*(1956)$ contribution and the theoretical spectrum [20]. It is clearly seen that agreement between the calculated and the experimental spectrum of the $pd\gamma$ reaction becomes significantly better [22].

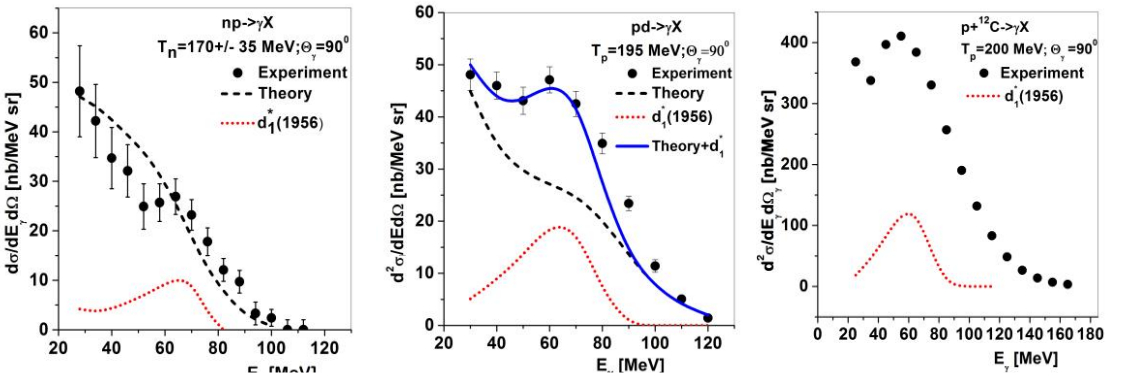

Fig.2.(Color online) The energy spectra of photons emitted at $90^0$ (full circles) from the reactions $np\to\gamma X$ at **170±35** MeV [15], $pd\to\gamma X$ at **195** MeV [17] and $p{}^{12}C\to\gamma X$ at 200 MeV [18]. The dashed lines are the theoretical predictions [16,20] and the dotted lines show the simulated $d_1^*(1956)$ contributions [22]. The solid line is the sum of the calculation [20] and $d_1^*(1956)$ contribution [22].

## 3.2. Indications in two-photon invariant mass spectra of the reactions.

**Reaction $pp\to\gamma\gamma pp$.** The invariant mass spectra of the photon pairs emitted from this reaction at energies

**1.36** and **1.2** GeV [23] were measured by the *CELSIUS-WASA Collaboration* at the CELSIUS ring using the WASA detector facility and the pellet hydrogen target. All four particles of this reaction were detected in coincidence. A surprising feature of these spectra is that they both contain pronounced resonant structures located around a mass of **280** MeV. The conservative estimates of the statistical significance $S$ of these structures by the formulae $S=N_S/(N_S+N_B)^{1/2}$ amount to **4.5σ** for the spectrum measured at **1.36** GeV and **3.2σ** at **1.2** GeV, where $N_S$ and $N_B$ are the numbers of the events attributed to a signal and a background, respectively. Thus, if these structures refer to the signature of the same mechanism of the reaction $pp \rightarrow pp\gamma\gamma$, then its total statistical significance is **5.5σ**. These resonance-like structures were regarded by the authors of Ref. [23] as a possible manifestation of the dynamical formation of the S-wave dipion resonance $\sigma$ in the *pp* collision with its subsequently decay into two photons. We proposed an alternative interpretation of their origin according to which these structures are due to the dibaryon mechanism of two-photon production in *pp* collisions $pp \rightarrow \gamma d_1^*(1956) \rightarrow pp\gamma\gamma$ [24]. It was found that the Monte Carlo simulated invariant mass spectra of photon pairs from this mechanism at **1.36** and **1.2** GeV (see Fig. 3) reasonably well reproduce the experimentally observed spectra in the vicinity of the resonance-like structures. This fact is a valid argument in favour of the $pp \rightarrow \gamma d_1^*(1956) \rightarrow pp\gamma\gamma$ mechanism existence.

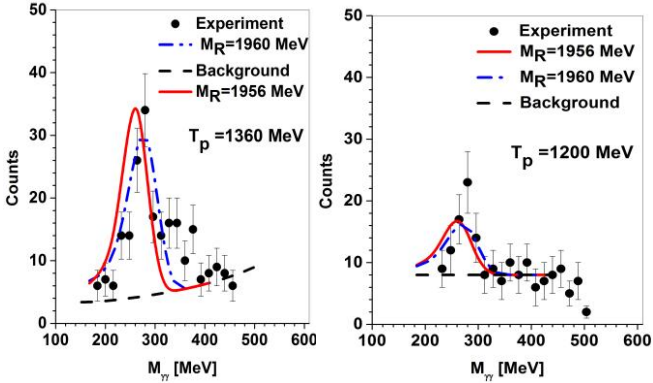

Fig.3. (Color online) Experimental two-photon invariant mass spectra of the reaction $pp \rightarrow pp\gamma\gamma$ (full circles) at **1.36** GeV (left) and 1.2 GeV (right) [23]. The dot-dashed and the solid lines show the Monte Carlo simulated two-photon invariant mass spectra of the process $pp \rightarrow \gamma d_1^* \rightarrow pp\gamma\gamma$ for the $d_1^*$ mass of 1956 MeV and 1960 MeV, respectively. The dashed line corresponds to the expected background due to the double $pp$-bremsstrahlung reaction.

**Reaction  p+$^{12}$C $\rightarrow \gamma\gamma$ X.** The two-photon invariant mass spectra of this reaction were measured at the proton synchrotron Saturne at Saclay, for several proton kinetic energies ranging from **800** MeV to **1500** MeV, by a two-arm neutral meson spectrometer **PINOT** [25]. In  Fig. 4 we show two such spectra borrowed from Ref. [26] and Ref. [27]. In these spectra the peaks corresponding to the $\pi^0$ and η mesons production are clearly seen. At the same time, one more peak can also be seen in both spectra. Our calculations show that this peak can be attributed to the dibaryon mechanism of photon production $NN \rightarrow \gamma d_1^*(1956) \rightarrow NN\gamma\gamma$.

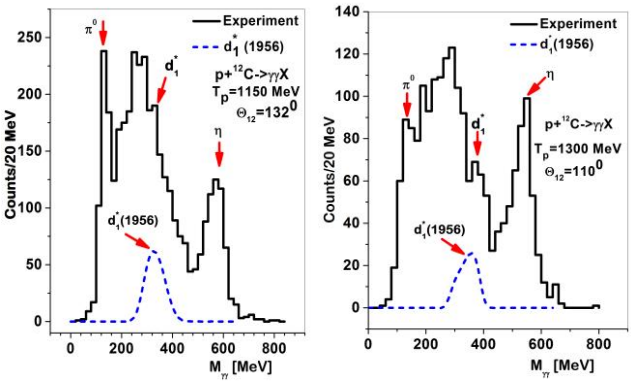

Fig. 4. (Color online) Experimental two-photon invariant mass spectra of the reaction $p+^{12}C \rightarrow \gamma\gamma X$ at **1150** MeV (left) [26] and **1300** MeV (right) [27]. The dashed lines correspond to the contribution of the dibaryon mechanism $pN \rightarrow \gamma d_1^*(1956) \rightarrow pN\gamma\gamma$.

**Reaction d+ $^{12}$C $\rightarrow \gamma\gamma$X.** The $\gamma\gamma$-invariant mass spectrum of this reaction was measured by a two-arm photon spectrometer at the incident deuteron momentum $p_d$ =2.75 GeV per nucleon [28]. A pronounced resonance structure at $M_{\gamma\gamma}$ =360 ± 7 ±9 MeV was observed in this spectrum. This spectrum along with the Monte Carlo simulated invariant mass spectrum of photon pairs from the process $NN \to \gamma d_1^*(1956) \to NN\gamma\gamma$ is shown in Fig.5. As can be seen, the calculated spectrum is in good agreement with the measured one. This fact strongly suggests that the origin of the structure is due to the $NN \to \gamma d_1^*(1956) \to NN\gamma\gamma$ process.

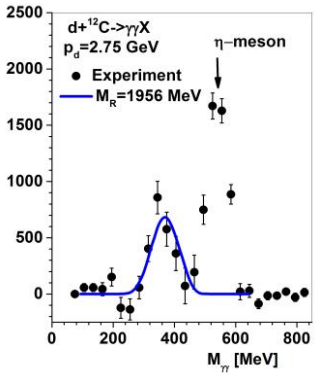

Fig. 5. (Color online) Experimental invariant mass spectrum of photon pairs from the reaction $d+{}^{12}C \to \gamma\gamma X$ at $p_d$ =2.75 GeV per nucleon (full circles) [28] compared to the Monte Carlo simulated invariant mass spectrum of photon pairs from the process $NN \to \gamma d_1^*(1956) \to NN\gamma\gamma$ (solid line) for the incident nucleon momentum $p_N$ =2.75 GeV.